\documentclass[longbibliography,floatfix,reprint,prb,showpacs,amsmath,amssymb,superscriptaddress]{revtex4-1}
\usepackage{graphicx}
\usepackage{amssymb}
\usepackage{color}
\usepackage{epstopdf}
\usepackage{bm}

\pdfoutput=1

\newcommand{\ve}{\mathbf}

\newcommand*\oline[1]{%
  \vbox{%
    \hrule height 0.5pt
    \kern0.25ex
    \hbox{%
      \kern-0.15em
      \ifmmode#1\else\ensuremath{#1}\fi
      \kern-0.1em
    }
  }
}

\begin{document}

\title{Giant spin orbit interaction due to rotating magnetic fields in graphene nanoribbons}

\author{Jelena Klinovaja}
\affiliation{Department of Physics, University of Basel,
             Klingelbergstrasse 82, CH-4056 Basel, Switzerland}
\author{Daniel Loss}
\affiliation{Department of Physics, University of Basel,
             Klingelbergstrasse 82, CH-4056 Basel, Switzerland}

\date{\today}

\begin{abstract}
We theoretically study graphene nanoribbons in the presence of spatially varying magnetic fields produced e.g. by nanomagnets. We show both analytically and numerically that an exceptionally large Rashba spin orbit interaction (SOI) of the order of $10$ meV can be produced by the non-uniform magnetic field. 
As a consequence, helical modes exist in armchair nanoribbons that exhibit nearly perfect spin polarization and are robust against boundary defects.
This paves the way to realizing spin filter devices in graphene nanoribbons in the temperature regime of a few Kelvins.
If a nanoribbon in the helical regime is in proximity contact to an $s$-wave superconductor, the nanoribbon can be tuned into a topological phase sustaining Majorana fermions.
\end{abstract}
\maketitle

\section{Introduction}

The last decade has seen remarkable progress in the physics and fabrication of graphene-based systems. \cite{Novoselov_2005, Novoselov_2009} The recent advances in producing graphene nanoribbons (GNRs) enable to assemble them with well-defined edges, in particular of armchair type. \cite{nanoribbon_production, nanoribbon_production_CNT} Moreover, it has been shown that the presence of adatoms can significantly increase the strength of the spin orbit interaction (SOI) of Rashba type. \cite{Rashba_2012} All this together makes nanoribbons  promising candidates for spintronics effects. In particular, generation of helical states, modes which transport opposite spins in opposite directions, is of great interest. Such modes were proposed in semiconducting nanowires, \cite{streda} 
carbon nanotubes, \cite{cnt_helical_2011, cnt_MF_2012} bilayer graphene, \cite{bilayer_MF_2011} and experimentally reported for quantum wires in GaAs hole gases. \cite{Quay_2010}
 They find applications in spin-filters, \cite{streda} Cooper pair splitters, \cite{tserkovnyak} and, in contact with an $s$-wave superconductor, they provide a platform for Majorana fermions  with non-abelian braiding statistics. \cite{Alicea_2012}

In the present work we propose a novel way to generate a giant effective SOI in GNRs by spatially varying magnetic fields that can be produced by nanomagnets. \cite{exp_field} This approach has an advantage over using adatoms because the surface of graphene is not in tunnel-contact with other atoms, which usually leads to high disorder with strong intervalley scattering. As we will see, large values of SOI result in helical modes of nearly perfect polarization. Moreover, nanoribbons, in stark contrast to semiconducting nanowires, have considerably larger subband splittings, allowing for a superior control of the number of propagating modes and of the gaps that are characteristic for the helical regime.

Further, our proposal is a next step in bringing topological features to graphene systems. Topological states proposed by Kane and Mele \cite{kane_mele} turned out  to be experimentally undetectable due to the small intrinsic SOI of graphene. In contast, we show here that if a GNR in the helical regime is brought into proximity to an s-wave superconductor, the system can be tuned into a topological phase that supports Majorana fermions. This opens up the possibility to use GNR for topological quantum computing.

The low-energy physics of armchair GNRs is characterized by broken valley degeneracy enforced by the boundary effects. \cite{brey_2006} To generate helical states we also need to lift the spin degeneracy. This can be achieved by magnetic fields in two ways: by a uniform magnetic field and Rashba SOI or by a spatially varying magnetic field. The chemical potential should be tuned inside the gap opened,
leading to a helical regime. We will study these two scenarios both analytically and numerically. Moreover, we will show numerically that the presence of helical states is robust against small non-idealities of the GNR edges.  This shows that  our proposal is realistic and experimentally feasible.

\begin{figure}[!b]
 \centering
 \includegraphics[width=\columnwidth]{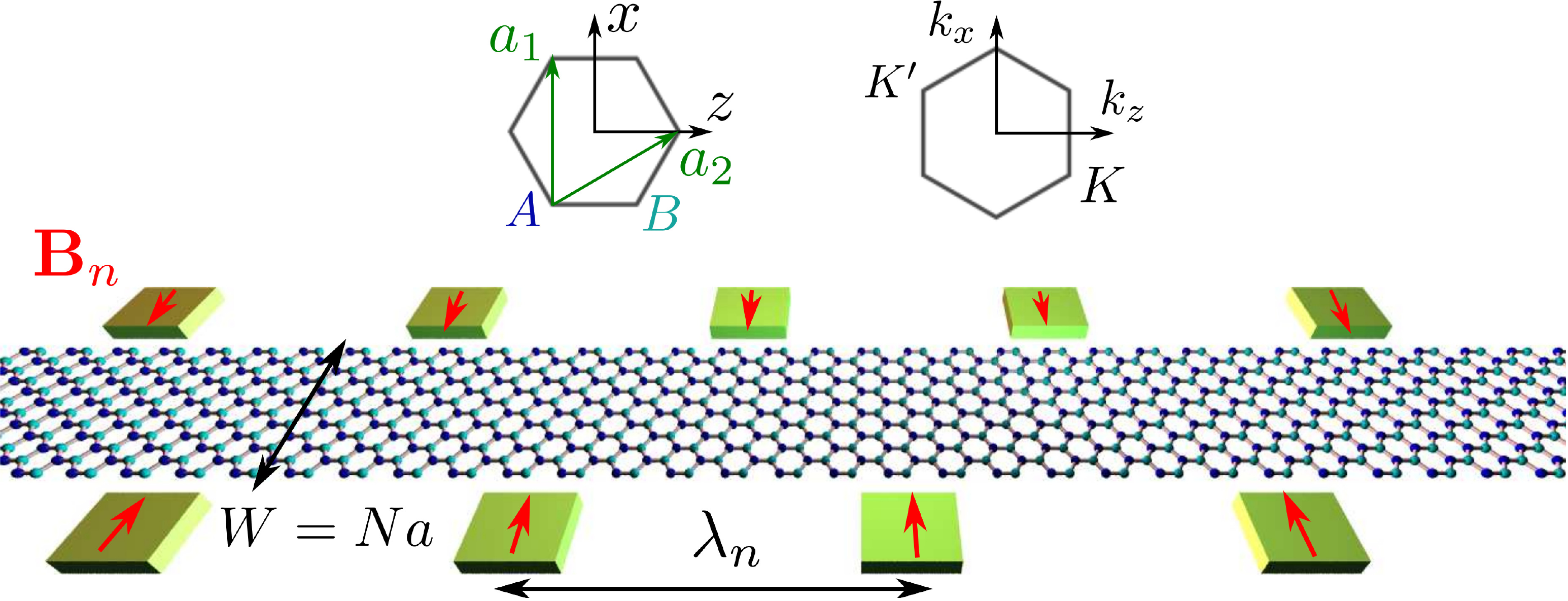}\\
 \caption{An armchair GNR formed by a finite strip of graphene aligned along the $z$-axis and of width $W$ in $x$-direction. The GNR is composed of two types of atoms $A$ (blue dot) and $B$ (green dot) and is characterized by hexagons in real space with translation vectors ${\bf{a}}_1$ and ${\bf{a}}_2$.  
The low-energy physics is determined by the momenta ${\bf k}=(k_x,k_z)$ around the two valleys ${\bf K}=-{\bf K}^\prime=(4\pi/3a, 0)$. Nanomagnets (green slabs) 
placed with period $\lambda_n$ on the sides of the GNR provide a spatially varying magnetic field ${\bf B}_n$ (red arrows).}
 \label{fig:system}
\end{figure}

\section{Graphene nanoribbon}
GNRs are strips of graphene,
a honeycomb lattice defined by translation vectors ${\bf{a}}_1$ and ${\bf{a}}_2$ and composed of two types of atoms, $A$ and $B$,  see Fig.~\ref{fig:system}. The GNR axis is chosen along the $z$-axis and has a finite width in $x$ direction.
GNRs are usually characterized by a width $W$ and a chiral angle $\theta$, the angle between the GNR axis and ${\bf{a}}_1$.  We only consider armchair nanoribbons for which $\theta$ is equal to $\pi/2$.

Graphene can be analyzed in the framework of the tight-binding approach. The effective Hamiltonian includes hoppings of electrons between neighboring sites,
\begin{equation}
\oline{H}_0=\sum_{<ij>,\lambda,\lambda'} t_{ij,\lambda\lambda'} c^\dagger_{i\lambda}c_{j\lambda'}.
\label{tight-binding}
\end{equation}
Here, $c_{i\lambda}$ are the standard electron operators, $i$ and $j$ are nearest-neighbor sites, and $\lambda, \lambda' $  are spin projections on the $z$-axis. Without SOI, the spin is  conserved and the hopping amplitude becomes $t_{ij,\lambda\lambda'}=t_{ij}\delta_{\lambda \lambda'}$, where $t_{ij}$ is spin-independent. It is more convenient to treat $\oline{H}_0$ in momentum space $(k_x, k_z)$.
The low-energy physics of graphene is determined by two valleys around ${\bf K}=-{\bf K}^\prime=(4\pi/3a, 0)$, where $a = |{\bf{a}}_1|$ is the lattice constant. Wavefunctions can be represented in the form 
$\psi=\sum_{\tau\sigma} \phi_{\sigma\tau} e^{i \tau K_x x}$, where $\tau=\pm1$ corresponds to $K/K^\prime$ and $\sigma=\pm1$ to the $A/B$  sublattice.
The Hamiltonian for the slowly-varying wavefuctions $\phi_{\sigma\tau}(x,z)$ is written in terms of the Pauli matrices $\sigma_i$ ($\tau_i$), acting on the sublattice (valley) degrees of freedom, as
\begin{equation}
H_0=\hbar \upsilon_F (\tau_3 k_x \sigma_1 +  k_z \sigma_2).
\label{graphene_zero}
\end{equation}
Here, $k_z$ ($k_x$) is the longitudinal (transverse) momentum calculated from a Dirac point, and $\upsilon_F$ is the Fermi velocity. 
From now on we work in the basis $\Phi=(\phi_{AK},\phi_{BK},\phi_{AK^\prime},\phi_{BK^\prime})$.

\begin{figure}[!t]
 \centering
 \includegraphics[width=\columnwidth]{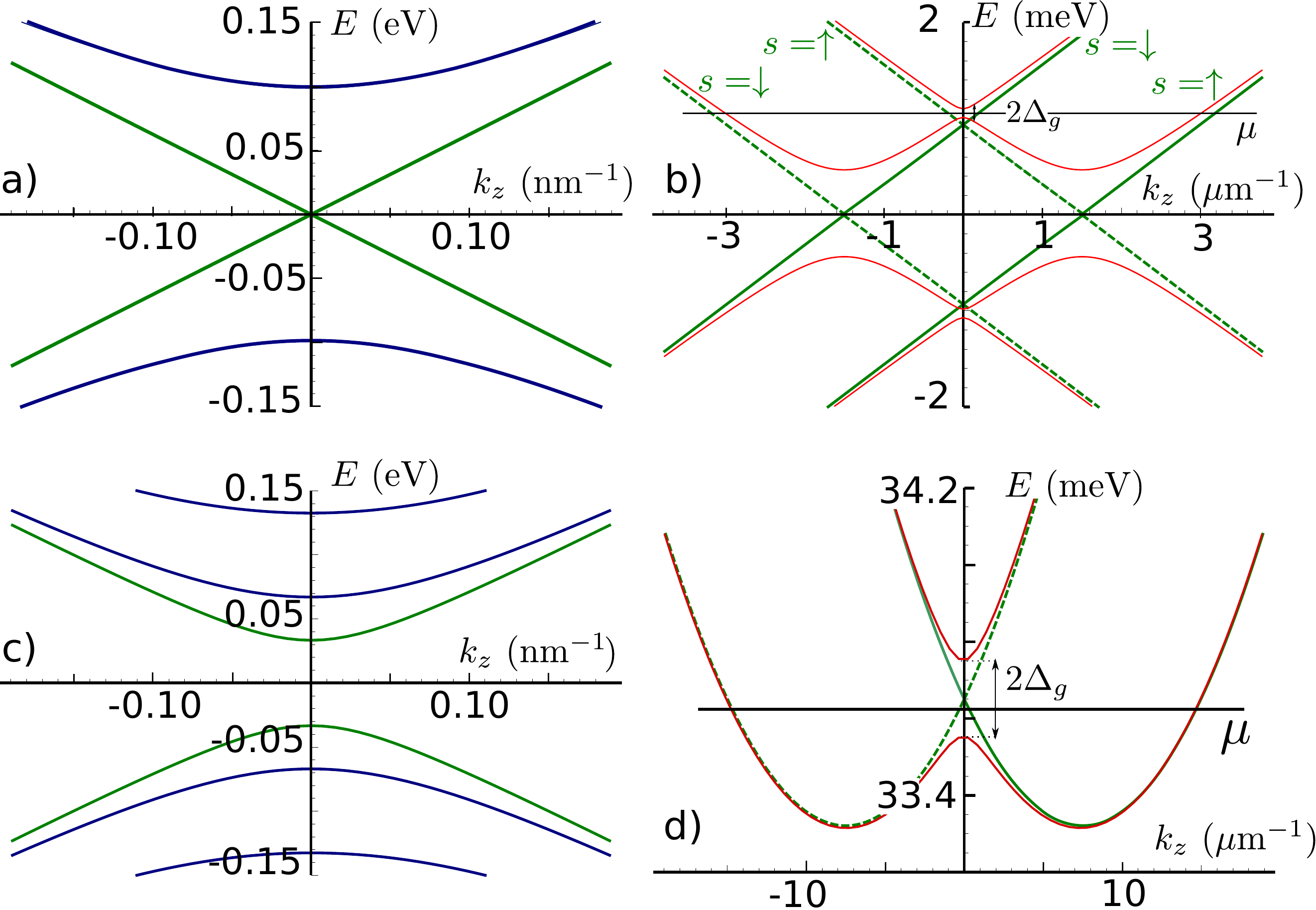}\\
 \caption{The spectrum of an armchair GNR obtained by numerical diagonalization of the tight-binding Hamiltonian $\oline{H}_0+\oline{H}_{so}+\oline{H}_z$.
The low-energy spectrum is linear for metallic [(a,b) with $N=82$] and quadratic for  semiconducting [(c,d) with $N=81$] GNRs. 
The SOI [(b) $\Delta_{so}=1\ \rm meV$ and (d) $\Delta_{so}=5\ \rm meV$] lifts the spin degeneracy, so the spectrum consists of (b) two Dirac cones or (c) two parabolas shifted by $k_{so}$ from zero (shown by green lines). For a metalllic GNR each branch is characterized not only by the spin projection $s$ but also by the isospin $\gamma$. The solid (dashed) lines correspond to $\gamma=1$ ($\gamma=-1$), see (b). 
While for a semiconducting GNR a magnetic field, $\Delta_Z=0.1\ \rm meV$, alone opens a gap $2\Delta_g$  [(d)], we also need to include intervalley scattering (modeled by fluctuations in on-site energies) for the metallic GNR [(b)]. 
If  the chemical potential $\mu$ is tuned inside the gap, the system is in the helical regime with nearly perfect polarization, $\langle s_x\rangle\approx 0.99$, in both cases.
}
 \label{fig:spectrum}
\end{figure}

A GNR, in contrast to a graphene sheet, is of finite width $W$ leading to well-gapped subbands. 
 In order to impose open boundary conditions on a GNR consisting of
$N$ unit cells
in transverse direction,
we effectively extend the GNR by two unit cells, so that the width is equal to $W^\prime=(N+2)a$ and impose vanishing boundary conditions on these virtual sites,
$\psi(0,z)=0$ and $\psi(W^\prime,z)=0$. \cite{brey_2006}
This leads to quantization of the transverse momentum $k_x$, 
$K_x+k_x = \pi n/W^\prime$,
where $n$ is an integer.

If the width of the GNR is such that $N=3M+~1$, where $M$ is a positive integer,
the GNR is metallic. \cite{brey_2006} The low-energy spectrum is linear and given by $E_{\sigma}=\gamma \hbar \upsilon_F k_z$, where
the isospin $\gamma\pm 1$, corresponding to the eigenvalues of the Pauli matrix $\sigma_2$.
The higher levels are two-fold degenerate (apart from spin, see below) and gapped by $\pi \hbar \upsilon_F/3(M+1)a$ (see  Fig.~\ref{fig:spectrum}a).

If the GNR width is such that $N=3M$  (or $N=3M+2$), where $M$ is a positive integer, the GNR is semiconducting. \cite{brey_2006} The gap at $k_z=0$ is given by $2 \hbar \upsilon_F |k_x^{min}|$, with $|k_x^{min}|=\pi/3(N+2)a$ .
In case of a semiconducting GNR all orbital states are non-degenerate (see Fig. \ref{fig:spectrum}c).
This means that the boundaries induce intervalley coupling and break the valley degeneracy \cite{falko, brey_2006,akhmerov}.

If the chemical potential $\mu$  
crosses only the lowest level of the  spectrum, there are two states propagating in opposite directions with opposite 
isospins $\sigma$. 
However, so far we have not taken spin into account, which will lead to four states at the Fermi level in total. As we will see next, this degeneracy can also be lifted if we include Rashba SOI and a uniform magnetic field or, equivalently, a spatially varying magnetic field.

\section{Rashba spin orbit interaction} 

The Rashba SOI arises from  breaking inversion symmetry. This can be caused  by an electric field $E_{ext}$ applied perpendicular to the GNR plane, or alternatively, by adatoms, which produce local electric fields. In the first case,  the SOI is quite small, $\Delta_{so} = eE_{ext} \xi $ with $\xi=4\times 10^{-5}\ \rm nm$ \cite{huertas_SOI,klinovaja_cnt} for realistic fields $E_{ext}\sim 1\ {\rm V}/ \mu{\rm m}$. In the second case, the strength of the SOI is significantly increased by doping, and values for $\Delta_{so}$ of $10 -100\ \rm {meV}$ have been observed
\cite{Rashba_2012}.
The general form of the Rashba SOI Hamiltonian can be obtained from symmetry considerations, \cite{kane_mele}
\begin{equation}
H_{so}=\Delta_{so} (\tau_3 s_z \sigma_1-s_x \sigma_2),
\label{so_symmetry}
\end{equation}
where the Pauli matrices $s_i$ act on the  spin. 

\section{Rotating magnetic field.} 
An alternative approach to generate helical modes is to apply a spatially varying magnetic field. \cite{Braunecker_Jap_Klin_2009,Flensberg_Rot_Field,Two_field_Klinovaja} Such a field can be produced by nanomagnets \cite{exp_field} or by vortices of a superconductor. We emphasize that this scheme requires not  perfect periodicity of the field but just a substantial weight of the Fourier component at twice the Fermi wavevector.   
Moreover, this mechanism is valid for both, rotating and linearly oscillating fields. 
For simplicity we assume in this section that the Rashba SOI discussed above is negligible. The case when both, a spatially varying magnetic field and Rashba SOI, are present was discussed recently in the context of  nanowires \cite{Two_field_Klinovaja}
and shown to lead to a number of striking effects such as 
fractionally charged fermions. \cite{Two_field_Klinovaja}

First, we consider a field rotating in a plane perpendicular to the GNR, leading to the Zeeman term
\begin{equation}
H_n^\perp=\Delta_Z\left[ s_y \cos (k_n z)+s_z \sin (k_n z)\right],
\end{equation}
where $\Delta_z = g \mu_B B/2$, $\mu_B$ is the Bohr magneton, $g$ the $g$-factor, and $\lambda_n = 2\pi/k_n$  the period of the rotating field.
It is convenient to analyze the position-dependent Hamiltonian $H_0+H_{n}^\perp$ in the spin-dependent rotating basis. \cite{Rashba_2003,Braunecker_Jap_Klin_2009} The unitary transformation $U_n= \exp(i k_n z s_x/2)$ brings the Hamiltonian back to a GNR with in-plane Rashba SOI and  uniform field  perpendicular to the GNR,
\begin{align}
H^\perp&=U_n^\dagger (H_0+H_n^\perp) U_n = H_0 + \Delta_Z s_y + \Delta_{so}^n s_x \sigma_2.
\end{align}
In a similar way, a field rotating in the plane of a GNR,
\begin{equation}
H_{n}^\parallel=\Delta_Z\left[  s_x \cos (k_n z)+s_z \sin (k_n z)\right],
\end{equation}
is equivalent to a GNR with out-of-plane Rashba SOI together with a uniform field along the $x$-axis,
\begin{align}
H^\parallel&=U_n^\dagger (H_0+H_n^\parallel) U_n
= H_0 + \Delta_Z s_x+\Delta_{so}^n s_y \sigma_2 .
\end{align}
The induced SOI favors the direction of spin perpendicular to the applied rotating magnetic field, and its strength is given 
by $\Delta_{so}^n=\hbar \upsilon_F k_n/2$,  independent of
the amplitude $\Delta_Z$.
For example, $\Delta_{so}^n$ is equal to $10\ \rm meV$ for nanomagnets placed with a period of $200\ \rm nm$.

\section{Helical modes.} The spectrum of $H^\perp$ (or by analogy of $H^\parallel$) can be easily found using perturbation theory. Taking into account that realistically $\Delta_Z \ll \Delta_{so}^n$, we treat the Zeeman term as a small perturbation. The induced SOI, given by $\Delta_{so}^n s_x \sigma_2$, leads to spin-dependent shifts of the $k_z$-momenta by $k_{so}=\Delta_{so}^n/\hbar \upsilon_F=k_n/2$, both for the metallic and the semiconducting GNRs, see Figs. \ref{fig:spectrum}b and \ref{fig:spectrum}d. Every level is characterized by the spin projection $s=\pm1$ on the $x$-axis, so the spin part of the wavefunctions, $|s\rangle$, is an eigenstate of the Pauli matrix $s_x$. The corresponding  spectrum and wavefunctions that satisfy the vanishing boundary conditions (for $\psi$) are given by 
\begin{align}
&\Phi^{E,k_z}_{\gamma,s} =e^{i z(k_z+sk_{so}) } (   -i \gamma, 1, i \gamma , -1) |s\rangle,
\label{met_wavefunc}\\
&E_{\gamma,s}= \gamma \hbar \upsilon_F (k_z + s k_{so})
\end{align}
for a metallic GNR and
\begin{align}
&\Phi^{E,k_z}_{\pm,s} =e^{i z(k_z+sk_{so}) } (\pm e^{i\varphi_s+i xk_x^{min} }, e^{i xk_x^{min} },\nonumber\\ 
&\ \ \ \ \ \ \ \ \ \ \   \ \ \ \ \ \ \ \ \ \ \ \  \ \  \mp e^{i\varphi_s-i xk_x^{min} }, 
 -e^{-i xk_x^{min} }) |s\rangle,
 \label{sem_wavefunc}\\
&E_{\pm,s}= \pm \hbar \upsilon_F \sqrt{ (k_x^{min})^2+(k_z + s k_{so})^2}
\end{align}
for a semiconducting GNR. Here we use the notation $e^{i\varphi_s}=[k_x^{min} - i  (k_z+sk_{so})]/\sqrt{(k_x^{min})^2 +(k_z+sk_{so})^2 }$.

A uniform magnetic field that is perpendicular to the spin-quantization axis defined by the SOI results in the opening of a gap $2\Delta_g$ at $k_z=0$. Using the wavefunctions given by Eq. (\ref{sem_wavefunc}), we can show that $\Delta_g =\Delta_Z k_x^{min}/\sqrt{(k_x^{min})^2+k_{so}^2}\approx \Delta_Z$
 for a semiconducting GNR. The spin polarization in this state is given by $|\langle s_x\rangle| \approx 1-(\Delta_Z k_x^{min}/4 \hbar \upsilon_F k_{so}^2)^2$.
In contrast to that, a metallic GNR possesses an additional symmetry.  Each branch is characterized not only by spin ($s=\pm1$) but also by  isospin ($\sigma=\pm1$), see Fig. \ref{fig:spectrum}b. Thus,
a magnetic field alone cannot 
 lift the degeneracy at $k_z=0$. However, if we include also terms breaking the sublattice symmetry, such as  intervalley scattering described by
$H_{KK^\prime}=\Delta_{KK^\prime}\tau_1$,
a gap will be opened.
Here, $\Delta_{KK^\prime}$ is the strength of the intervalley scattering, which can be caused by impurities or fluctuations in the on-site potential. Assuming $\Delta_Z,\Delta_{KK^\prime}\ll\Delta_{so}$, the  gap becomes  $2\Delta_g=2 \Delta_{KK^\prime} \Delta_Z/\Delta_{so}$ in leading order. The spin polarization of the helical states is given by $|\langle s_x\rangle| \approx 1-(\Delta_Z/\Delta_{so})^2$. We note that for both semiconducting and metallic GNR, $\Delta_Z$ limits the size of the gap $\Delta_g$.

We note that $H^\perp$ is equivalent to the Hamiltonian describing a GNR in the presence of Rashba SOI and a uniform magnetic field applied in perpendicular $y$-direction,  $H_{tot}=H_0+H_{so}+H_Z$ [see Eqs. (\ref{graphene_zero}) and (\ref{so_symmetry})] in first order perturbation theory in the SOI. Here, the Zeeman term is given by $H_Z = \Delta_Z s_y$.
The wavefunctions given by Eqs. (\ref{met_wavefunc}) and (\ref{sem_wavefunc}) are eigenstates of the Pauli matrix $\tau_1$,
 so the diagonal matrix element of $\tau_3$ is zero. This leads to the result that the term $\tau_3 s_z \sigma_1$ in the Rashba Hamiltonian $H_{so}$ averages out in first order perturbation theory, and $H_{tot}$ is indeed equivalent to $H^\perp$.
This means that the effect of the SOI
is a spin-dependent shift of $k_z$ by $k_{so}=\Delta_{so}/\hbar \upsilon_F$. Similarly, the uniform magnetic field opens a gap at $k_z=0$, which can be as big as $10 \ \rm K$ for a field of about $10 \ \rm T$. 

An alternative approach to above perturbation theory  is to analyze the GNR with Rashba SOI analytically. For graphene the spectrum of the effective Hamiltonian $H_0+H_{so}$ is given by
$
E_{j,\pm}=\pm \left(\Delta_{so} + j \sqrt{(\hbar \upsilon_F k_x)^2 +(\hbar \upsilon_F k_z)^2+ \Delta_{so}^2}\right)
$,
where the index $j$ is equal to $1$ ($-1$) for the highest (lowest) electron level, and the $\pm$ sign distinguishes between electrons and holes.
The SOI lifts the spin degeneracy, however, the valley degeneracy is maintained, and $\tau_3$ is a good quantum number. 
Analogously to Ref. \cite{brey_2006}, we search for a sum over the eigenstates $\psi_{\tau,q} (x)$ of $H_0+H_{so}$,
$
\psi^{E,k_z}(x,z)= \sum_{\tau,q} b_{\tau,q} \psi^{E,k_z}_{\tau,q} (x),
$
such that the boundary conditions are satisfied.
The index $q=(j,\pm)$ distinguishes between four wavevectors satisfying $E_{j, \pm}(k_x=\pm k_{j})=E$,
$\hbar \upsilon_F k_{1,2}=\sqrt{E^2 - (\hbar \upsilon_Fk_z)^2 \pm 2E \Delta_{so}}\label{k_1}$.
We also introduce new variables $\theta$ and $\gamma$, via
$\cos \theta = {\hbar \upsilon_F k_z}/{E}$ and $\sqrt{2} \sin \theta \sin \gamma ={\hbar \upsilon_F k_1}/{E}$.
We allow for real as well as imaginary values of $k_{1,2}$, $\theta$, and $\gamma$.
The spectrum of a metallic GNR is then given implicitly by
\begin{widetext}
\begin{align}
\tan^2 \theta \left(\sin\left(\frac{k_1 W}{2}\right)\cos\left(\frac{k_2 W}{2}\right)+\cos\left(\frac{k_1 W}{2}\right)\sin\left(\frac{k_2 W}{2}\right)\sin (2\gamma)\right)^2=-\sin\left(k_1 W\right)\sin\left(k_2 W\right)\sin (2\gamma).
\label{analytical_solution}
\end{align}
\end{widetext}
The exact solution defined by Eq. (\ref{analytical_solution}) can be analyzed analytically by means of Tailor expansion. For example,
if $\Delta_{so}\ll \hbar \upsilon_F k_z$, we get $E=\pm \hbar \upsilon_F k_z\pm \Delta_{so}$, which is in agreement with previous perturbative calculations.

\section{Numerics} To check our analytical results numerically, we extend the tight-binding Hamiltonian $\oline{H}_0$ by allowing for hoppings with spin-flip,
\begin{equation}
\oline{H}_{so}=\sum_{<ij>, \lambda, \lambda'} i c_{i\lambda}^\dagger {\bf u}_{ij}\cdot {\ve s}_{\lambda \lambda'} c_{j\lambda'}+ {\rm H.c.},
\label{tb_soi}
\end{equation}
in such a way that $\oline{H}_{so}$ is equivalent to the Rashba SOI in the low-energy sector. 
Here, ${\ve s}_{\lambda \lambda'}$ is a vector composed of the Pauli matrices, and spin-dependent hopping elements are defined as
${\bf u}_{ij}=-(3 \Delta_{so}/4) {\bf z}\times {\boldsymbol e}_{ij}$.
A unit vector ${\boldsymbol e}_{ij}$ points along the bond between two sites $i$ and $j$. 
The results of the numerical diagonalization of the Hamiltonian $\oline{H}_0+\oline{H}_{so}+\oline{H}_Z$ are presented in Fig. \ref{fig:spectrum}, where the Zeeman term corresponding to a magnetic field $\ve B$ is modeled as
\begin{equation}
\oline{H}_Z = \sum_{i, \lambda, \lambda'}  c_{i\lambda}^\dagger {\ve B } \cdot {\ve s}_{\lambda \lambda'} c_{i\lambda'}.
\end{equation}
As shown in Fig. \ref{fig:spectrum}, the numerical results fully confirm the analytical calculations. 

\section{Stability against edge defects.} 
\begin{figure}[!t]
 \centering
 \includegraphics[width=\columnwidth]{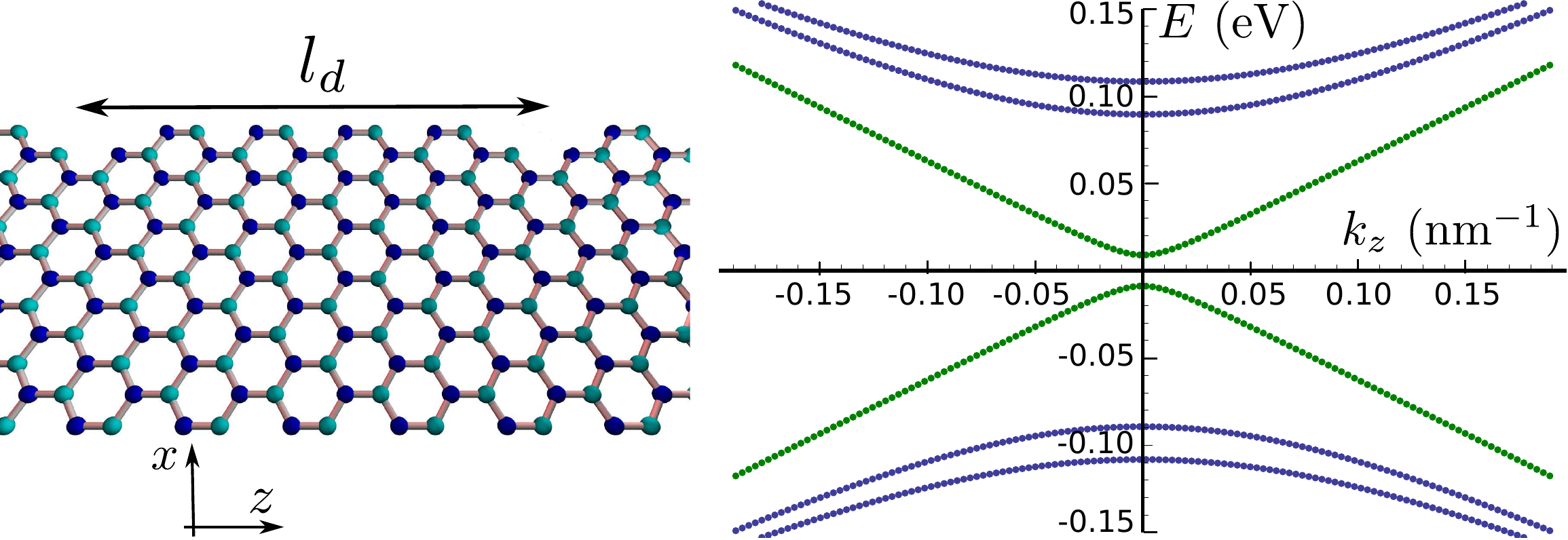}\\
 \caption{Defects on the edges of a metallic armchair GNR ($N=82$) result in opening of a gap at zero energy. In the numerical diagonalization, the defects were modeled by omitting two atoms on the edges, which was assumed to be periodic with period $l_d=5\sqrt{3}a$. We see that the spectrum changes only little and the qualitative features of  a metallic armchair GNR are maintained.
}
 \label{fig:edges}
\end{figure}

The spectrum of GNRs is known to be sensitive to the specific form of the edges. For example, the linear spectrum of a metallic GNR becomes parabolic for non-ideal armchair boundaries (see Fig. \ref{fig:edges}). In contrast to that, subband gaps are only slightly modified for semiconducting GNRs. 
We conclude that the valley degeneracy, in general, is lifted due to strong intervalley mixing induced by the boundaries and this is a  property of all armchair GNRs .\cite{akhmerov, falko}
We emphasize that for the scenario of helical modes developed above we do not need any specific symmetries. Thus, our proposal is robust against edge defects.

The scenario with a rotating magnetic field is even more universal. The only criterion is that the Fermi wavevector $k_F$ is not too large, typically $k_F/K_x$ should be smaller than $10^{-2}$. This is a natural limit resulting from the fact that the period of rotation of a magnetic field should be much larger than the lattice constant.

\section{Majorana fermions}
Next, we consider GNRs in the helical regime brought near to a superconductor.
If $\mu$ is tuned inside the gap opened by the field, the two propagating modes are helical.  
The proximity-induced superconductivity in the GNR leads to the coupling between such states and gaps the spectrum. The system can be effectively described in the linearized model for the exterior ($\chi=e$, states with momenta close to the Fermi momentum, $k_e=k_F$) and the interior branches ($\chi=i$, states with nearly zero momenta, $k_i=0$). \cite{MF_wavefunction_klinovaja_2012} The electron operator is represented as
$\Psi(z)=\sum_{\rho={\pm 1},\chi=e,i} e^{i \rho k_\chi z} \Psi_{\rho\chi }$,
where the sum runs over the right ($R$, $\rho=1$) and left ($L$, $\rho~=~-1$) movers. 
The effective Hamiltonian becomes
\begin{align} 
H = -i \hbar \upsilon_F  \rho_3 \chi_3 \partial_x  + \frac{\Delta_g}{4}  \eta_3 \rho_1 (1+\chi_3) + \Delta_s  \eta_2 \rho_2,
\label{static}
\end{align}
where the Pauli matrices $\chi_i$ ($\eta_i$) act in the interior-exterior branch (electron-hole) space, and
$\widetilde{\Psi}=(\Psi_{Re},\Psi_{Le},\Psi_{Re}^\dagger,\Psi_{Le}^\dagger,\Psi_{Li},\Psi_{Ri},\Psi^\dagger_{Li},\Psi^\dagger_{Ri})$. 
Following Refs. \cite{MF_wavefunction_klinovaja_2012, Two_field_Klinovaja}, we find that the  criterion for the topological phase transition is given by $\Delta_g>\sqrt{\mu^2 + \Delta_s^2}$. In terms of Zeeman energy this gives  $\Delta_Z>\Delta_{so}\sqrt{\mu^2 + \Delta_s^2} /\Delta_{KK^\prime}$ ($\Delta_Z>\sqrt{\mu^2 + \Delta_s^2}\sqrt{(k_x^{min})^2+k_{so}^2} /k_x^{min}$) for a metallic (semiconducting) GNR.

\section{Conclusions} We have shown that helical modes can be generated in graphene nanoribbons by a spatially varying magnetic field or by Rashba spin orbit interaction with a uniform magnetic field. We have demonstrated that the opening of the gap is universal for both semiconducting and metallic graphene armchair nanoribbons independent of the mechanism that induces the spin orbit interaction, leading to a helical regime with nearly perfect spin polarization.  Moreover, we have checked numerically that the helical regime is robust against boundary defects. All this makes graphene nanoribbons  promising candidates for spin effects and spintronics applications.

\acknowledgments
This work is supported by the Swiss NSF, NCCR Nanoscience, and NCCR QSIT.

\end{document}